\documentclass[aps,prl,reprint,superscriptaddress]{revtex4-2}

\usepackage{amsmath,amssymb,graphicx}
\usepackage{hyperref}

\begin{document}

\title{Black Hole–Inspired Horizon Model for Neural Signal Dynamics}

\author{E. Canessa}
\affiliation{International Centre for Theoretical Physics, Trieste, Italy}


\begin{abstract}
Electroencephalographic (EEG) signals provide macroscopic observables
of complex neural dynamics. We introduce a horizon-inspired framework
in which measured EEG signals are modeled as projections of a complex
wave-like representation constrained by an effective boundary analogous
to an event horizon. In this formulation the signal amplitude obeys a
renormalization-group scaling relation while EEG spectral entropy
parameterizes the accessibility of observable modes. The resulting
solutions generate oscillatory structures whose geometry and spectral
signatures can be explored through signal analysis and sonification.
This mapping between entropy-based neural observables and wave-like
signal representations provides a physically motivated framework
linking entropy measures, scale-dependent dynamics, and observable
neural oscillations, and suggests testable connections between
spectral entropy and the amplitude scaling of EEG modes.
\end{abstract}

\maketitle

\emph{Introduction}--Electroencephalography (EEG) provides a noninvasive
measurement of neural dynamics through electrical signals recorded at the scalp. These signals exhibit complex
oscillatory patterns reflecting collective neuronal activity
across multiple temporal scales.
Although these signals offer direct experimental access to
brain activity, they represent only a limited projection of the underlying
neural processes \cite{buzsaki2006}. From a physics perspective, this
situation resembles systems in which observable quantities arise from
effective boundary descriptions while internal degrees of freedom remain
hidden.

Examples of such boundary phenomena occur in gravitational systems,
where observables accessible to external observers are determined
by processes occurring near an event horizon while the internal
structure remains inaccessible \cite{hawking1975,bekenstein1973}. 
Similar ideas arise in statistical physics and complex systems,
where macroscopic observables provide coarse-grained descriptions of
systems composed of many microscopic degrees of freedom
\cite{Anderson1972,Goldenfeld1999}.
This analogy motivates the exploration of whether
neural signals may admit an effective description in which EEG observables
correspond to boundary-level projections of deeper neural dynamics.
At the same time many complex systems are successfully described through scale-dependent effective theories governed by renormalization-group (RG) dynamics \cite{wilson1975}.

\begin{figure}[t]
\centering
\includegraphics[width=0.95\columnwidth]{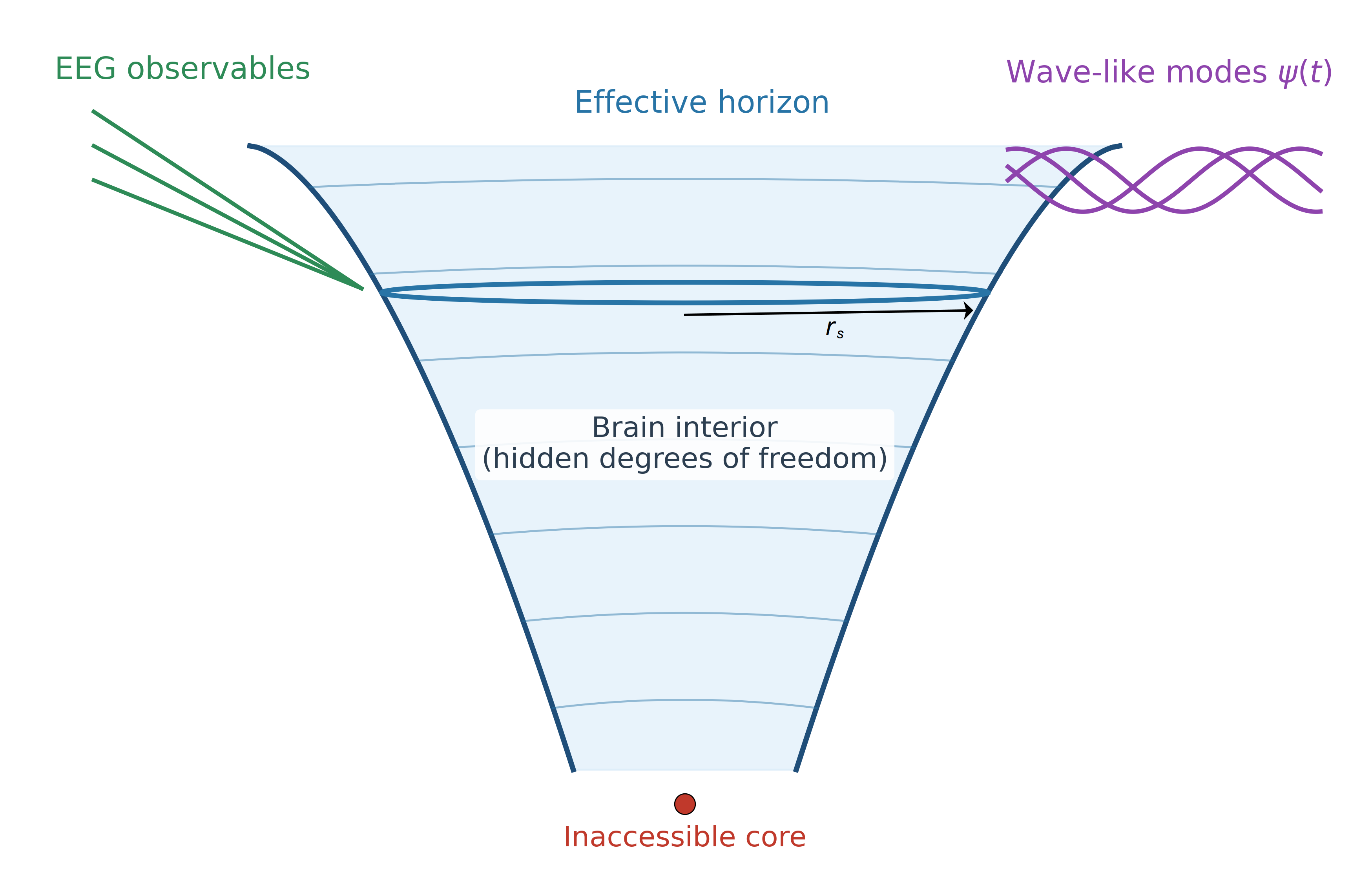}
\caption{
Schematic representation of the horizon-inspired model. Externally measured EEG observables correspond 
to signals accessible outside an effective horizon characterized
by the accessibility parameter $\Gamma_{r}$, measuring the distance from the effective
boundary $r_{s}$. Internal neural processes
remain hidden inside the boundary with an inaccessible core. Observable
signals emerge as wave-like modes $\psi(t)$ that can
be analyzed and sonified.
}
\label{fig:horizon}
\end{figure}

In this Letter we introduce a horizon-inspired representation of neural
signals in which EEG observables emerge as wave-like modes governed by
RG scaling and parameterized by spectral entropy. A
schematic representation of the horizon-inspired model is shown in Fig.~\ref{fig:horizon}.
The brain is modeled through an effective horizon-like boundary that
provides a formal analogy for observer-limited access to internal dynamics.
Within this framework, EEG
features associated with distinct mental states--i.e., varying
levels of awareness--are related to underlying brain wavefunctions. 
Real-valued EEG signals can then be rendered audible within the model.

\emph{The model}--We introduce a complex representation $\psi(r)$ describing neural
activity as a function of an abstract radial coordinate $r$.
The amplitude is assumed to evolve according to a RG transformation of the
form
\begin{equation}
r\frac{d|\psi(r)|}{dr} =
\beta_1 |\psi(r)| + \beta_3 |\psi(r)|^3 + \cdots \;,
\end{equation}
where $\beta_1$ and $\beta_3$ are real coefficients describing
scale-dependent behavior and $0 < r_{0} < r < \infty$.
To leading nonlinear order, the solution for the probability density
$\rho(r)=|\psi(r)|^2$ becomes
\begin{equation}
\rho(r) =
\frac{\beta_{1} (r/r_0)^{2\beta_{1}}}
{\beta_{3} [1 - (r/r_0)^{2\beta_{1}}] } \; ,
\end{equation}
assuming $\beta_{1}/\beta_{3} < 0$ for stability.
After normalization, the corresponding complex wavefunction in position space is then,
\begin{equation}\label{eq:psi}
\psi(r)= e^{i\theta(r)}
\sqrt{\frac{-\beta_1/\beta_3}{1-(r_{0}/r)^{2\beta_1}}} \; ,
\end{equation}
with an arbitrary phase factor represented by $\theta(r)$.
For $r \gg r_0$, the amplitude approaches a constant value
$|\psi|^{2} \rightarrow |\beta_{1}/\beta_{3}|$, analogous to RG fixed-point,
scale-invariant behavior observed in critical systems.

To connect the model with neural observables we define an accessibility index
\begin{equation}\label{eq:eq4}
\Gamma_{r} = 1 - \frac{r_s}{r} \;,
\end{equation}
representing the effective distance from a horizon-like boundary $r_{0} \to r_{s}$. 
Through this geometric construction, the quantity $\Gamma_{r}$ 
can be interpreted as a dimensionless
accessibility factor that quantifies how strongly internal
dynamics are filtered before becoming observable signals.
Near the horizon scale ($r \approx r_{s}$) the value $\Gamma_{r} \to 0$,
indicating that observable quantities are strongly
constrained by the boundary. Far from the horizon
($r \gg r_{s}$) the parameter approaches $\Gamma_{r} \to 1$, corresponding
to a regime in which internal configurations are more
fully accessible to external observation. In this sense
$\Gamma_{r}$ plays a role analogous to a ``redshift'' or transmission
factor linking internal dynamics to observable signals.

In the analogy adopted, the value $r = r_{s}$ represents an
effective horizon scale analogous to the Schwarzschild horizon
in gravitational systems.
Inside $r < r_{s}$, Eq.~\ref{eq:eq4} leads to a mathematical singularity
characterized by divergent curvature. Motivated by this structure, we propose a speculative association
between a Schwarzschild-like limit--representing irreversible spacetime
transitions where information cannot return--and a model projection
intended to capture aspects of mental thresholds. 
Using the accessibility parameter $\Gamma_{r}$ and fixing the constants
$2\beta_{1}=1$ and $2\beta_{3} \equiv -1/\beta < 0$, the wavefunction amplitude
of Eq.~(\ref{eq:psi}) can be rewritten in the form
\begin{equation}\label{eq:cossin}
|\psi(r)| \propto \sqrt{\frac{\beta}{\Gamma_{r}}} \; .
\end{equation}

Using the Bekenstein–Hawking relation $S=k_{B} A /(4L_{p}^{2})$
for a Schwarzschild horizon with area $A=4\pi r_{s}^2$, and the definition of
Eq.(\ref{eq:eq4}) for $\Gamma_{r}$ and the radial coordinate $r$, the entropy becomes
\begin{equation}
S^{*} \equiv \frac{S}{\pi k_{_{B}}} = \left( \frac{r_{s}}{L_{p}} \right)^{2} =
\left[ \left( \frac{r}{L_{p}} \right) \left( 1 - \Gamma_{r} \right) \right]^{2} > 0  \; .
\end{equation}
The Planck-like length $L_{p}$ enters the description for a dimensionless analysis (see details in \cite{Canessa26}).
This expression applies to regions with $r > r_{s}$. 
In statistical physics entropy measures the number of
accessible configurations. Within the present framework,
the accessibility parameter $\Gamma_{r}$ increases with the
effective distance from the horizon boundary. Therefore
larger entropy corresponds to a larger number of
accessible neural configurations and to a larger
effective distance from the horizon scale.

The phase angle $\theta(r)$ of the wavefunction determines the oscillatory structure of the model and, in turn, the audible projections. In the present framework, the phase is taken to be
\begin{equation}\label{eq:eq20}
\theta(r) = kr + \phi(r)  \; ,
\end{equation}
in terms of plane-wave contributions plus a function specific to the present system. This 
function is assumed to follow another simple RG-like equation
\begin{equation}\label{eq:eq21}
r\, \frac{d\phi(r)}{dr} = \beta_{2}\, |\psi(r)|^{2} + \cdots \; .
\end{equation}
Using Eq.~(\ref{eq:cossin}), the radial solution up to $\mathcal{O}(2)$ yields
\begin{equation}
\theta(r) = kr + \beta\, \beta_{2} \, \ln|r - r_{s}| + \phi_{_{0}}  \; ,
\end{equation}
which introduces logarithmic phase modulation near the horizon scale.
It is interesting to note that this assumption for $\theta(r)$ relates to the tortoise coordinate
$r_{*} = r + r_{s} \, \ln |r/r_{s} - 1|$,
which removes the Schwarzschild-coordinate singularity at $r=r_{s}$ and leads to plane-wave solutions
$\psi(r_{*},t) \sim e^{-iw (t\pm r_{*})}$.

\emph{Connection with EEG Observations}--To map features of the wavefunction $\psi(r)$ to audible
parameters--and to explore how it evolves--it is necessary to specify
how the real (or imaginary) parts of $\psi(r)$ generate waveforms suitable
for sound. First, to evaluate $\Re[\psi(r)]$, the theoretical Schwarzschild-like
entropy is identified with empirical measures of spectral entropy $S_{eeg}$
derived from spontaneous EEG measurements \citep{EEGLab}. 
Second, it is necessary to specify how time $t$ relates an
analogous spatial coordinate in order to evaluate $\Gamma_{r}$ within the present
construction. This isomorphism is obtained by mapping $r = v_{0} t$, 
where $v_{0}$ is a scaling parameter linking the
abstract coordinate to the temporal evolution of EEG measurements.

\begin{figure*}[t]
\centering
\includegraphics[width=0.48\textwidth]{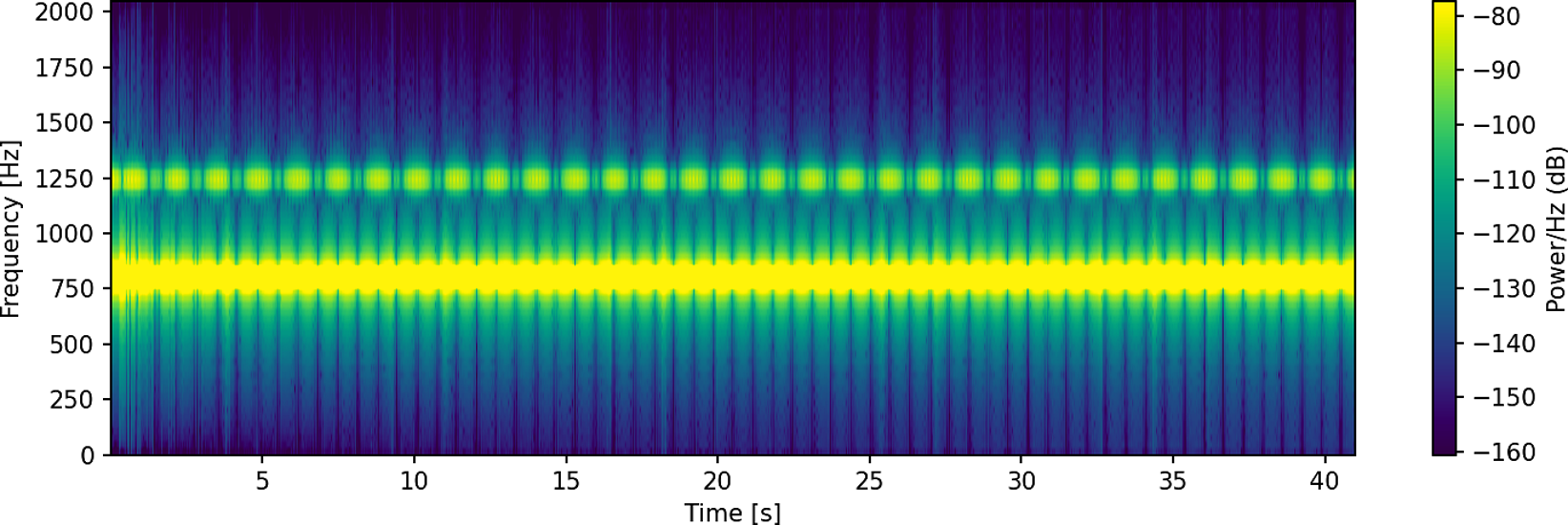}
\includegraphics[width=0.48\textwidth]{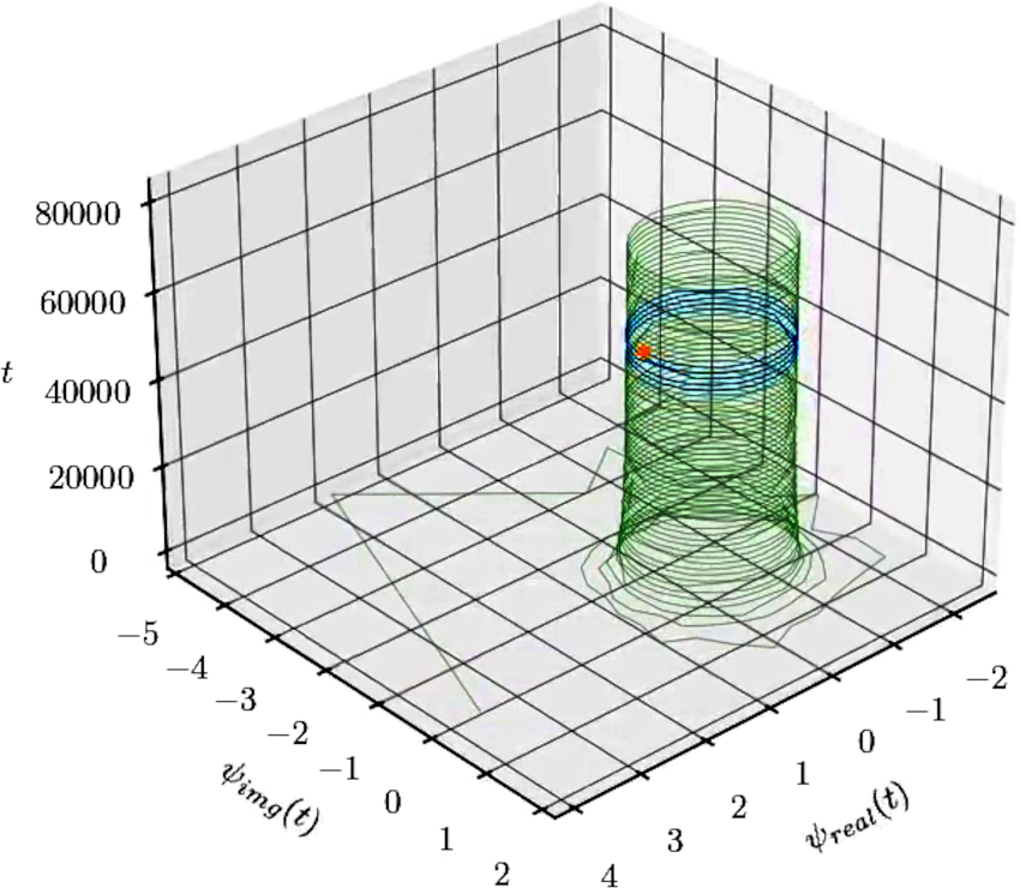}
\includegraphics[width=0.48\textwidth]{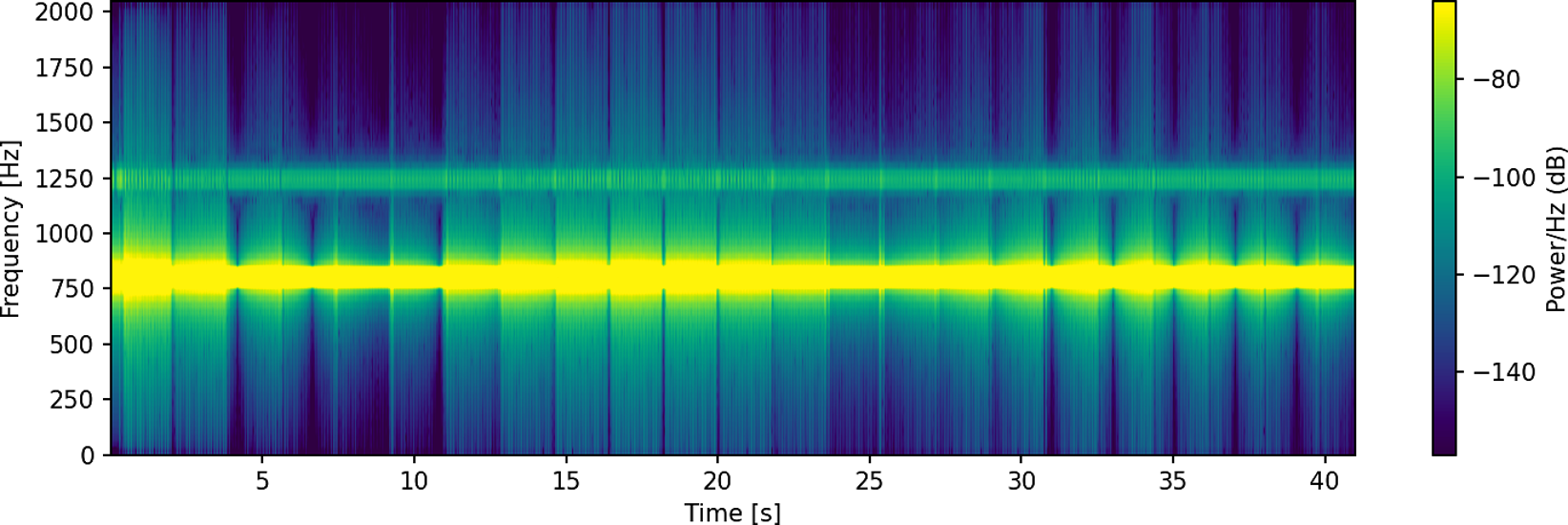}
\includegraphics[width=0.48\textwidth]{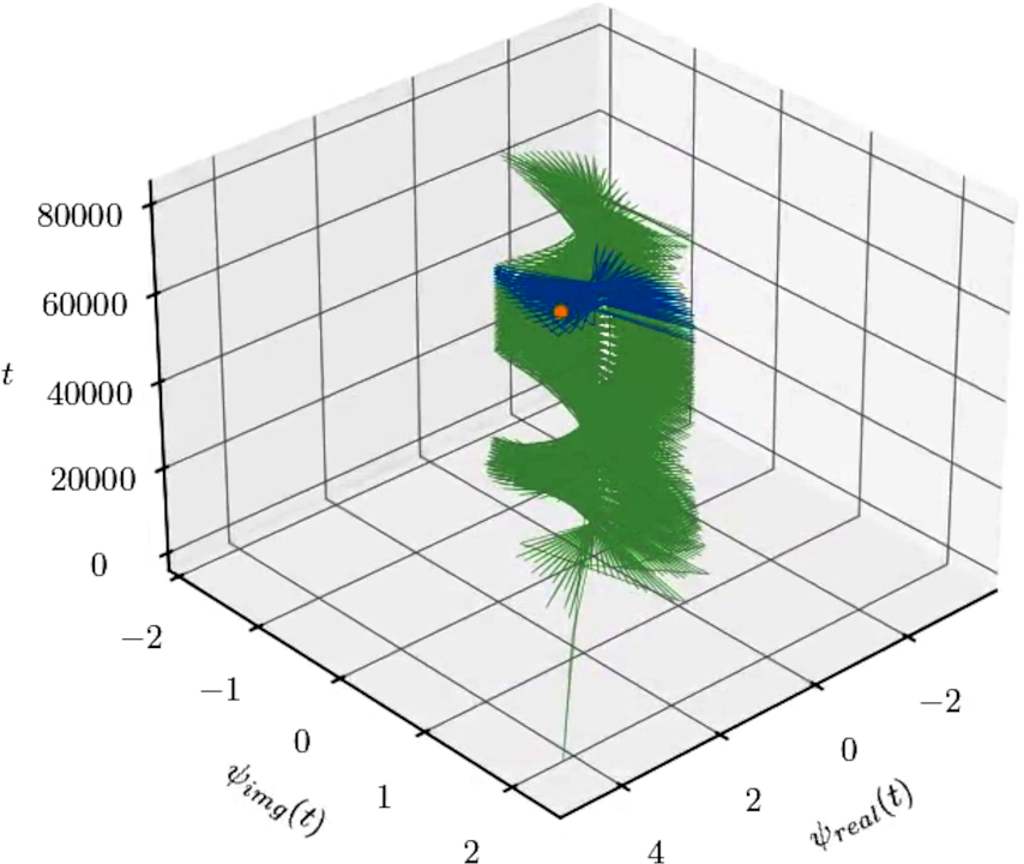}
\caption{
Wavefunction trajectories and corresponding spectrograms produced by the model using EEG recordings
in a female skull (Fpz-Cz channel) \citep{EEGLab}. Top panels show single-helical trajectories 
of the complex wavefunction while bottom
panels show double-helix regime obtained for larger angular frequency $\omega$.
}
\label{fig:sonification}
\end{figure*}

Substituting the time mapping yields the observable real part of the waveform 
\begin{equation}
\Re[\psi(t)] =
\sqrt{\frac{\beta}{\Gamma_{r}}}
\cos\left(\omega t + \beta \ln |t/t_{0} -1|^{\beta_{2}} + \phi_{1}\right) \; ,
\end{equation}
with $\omega\equiv k\,v_{_{0}}$, $t_{_{0}}\equiv r_{s}/v_{_{0}}$ and
$\phi_{1} \equiv \phi_{_{0}} + \beta\, \beta_{2} \ln r_{s}$.
Here $t$ is treated as the time coordinate used by the external observer, aligned with the measurement of EEG events.
The complex representation provides a geometric interpretation of neural
signals and forms a family of oscillatory modes
whose spectral characteristics depend on entropy-derived
parameters. The real and imaginary components define trajectories in phase
space forming helical structures under time evolution whose projections correspond to
observable oscillatory EEG signals as shown in Fig.~\ref{fig:sonification}.

The RG equation governs the amplitude of this
wavefunction, yielding scale-invariant solutions whose observable structure can be parameterized by
EEG spectral entropy. The resulting modes generate characteristic oscillatory patterns that can be
rendered acoustically through sonification techniques \cite{can2022}.
Because the real component of $\psi(t)$ corresponds to an oscillatory
waveform, the representation naturally lends itself to sonification.
Sonification converts data into sound in order to reveal temporal
patterns that may not be immediately visible in standard plots.

An appealing feature of the wave-like representation
is that its real component can be interpreted directly as
an audio waveform. This allows neural dynamics to be
explored through sonification, transforming EEG signals
into audible structures that reveal temporal patterns in
a different perceptual modality.
To produce continuous signals from discrete EEG data
it is necessary to reconstruct the time series using Gaussian smoothing.
The reconstructed amplitudes are then normalized and
converted into digital audio samples. Spectrogram analysis of the resulting signals reveals structures that depend
on the parameters controlling the wavefunction representation. In particular, different parameter choices lead to
distinct helical trajectories in complex phase space whose
projections produce characteristic spectral patterns in
the audio domain.

\emph{Discussion}--The theoretical framework inspired by
concepts from horizon physics does not
assume that neural processes are quantum mechanical
or that the brain behaves as a gravitational system. Instead, mathematical structures drawn from quantum theory and horizon physics are used as organizing tools for
describing observable neural signals in terms of wave-like
representations. Within this interpretation $\Gamma_{r}$ acts as an effective
accessibility coordinate that links entropy, observable
signal amplitude, and the distance from the horizon scale.

The horizon-inspired representation leads to several qualitative
features that provide insight into the structure of neural signals.
First, the RG equation governing the amplitude introduces a natural
scale hierarchy in which the observable signal approaches a constant
value far from the effective boundary. In the context of EEG dynamics,
this behavior can be interpreted as the emergence of stable macroscopic
oscillations from a large ensemble of interacting neural units.
Second, the entropy dependence of the accessibility parameter suggests
that variations in neural complexity directly influence the amplitude
and phase structure of the observable modes. Higher spectral entropy
corresponds to a broader distribution of frequencies and therefore to
a richer set of accessible neural configurations. Within the present
framework this corresponds to a larger effective distance from the
horizon boundary of radius $r_{0} \to r_{s}$ and to a modification of the amplitude scaling of the
wave-like modes. In fact, transitions between neural states, such as different
sleep stages, could correspond to changes in the effective
accessibility parameter.
Third, the logarithmic phase term appearing in the model produces
oscillatory patterns that can exhibit log-periodic modulation. Such
structures are known to arise in systems governed by scale-invariant
dynamics and RG flows \cite{Sornette}. In the context of neural
signals this suggests that certain EEG patterns may reflect underlying
scale-dependent organization of neural activity.

The model also provides a geometric interpretation of neural signals in
terms of trajectories in complex phase space. Depending on parameter
values, these trajectories may form helical structures whose projections
onto real observables correspond to oscillatory EEG waveforms. This
representation offers an intuitive way to connect spectral properties
of neural signals with geometric structures in the underlying complex
representation.
Finally, the possibility of sonifying the wavefunction modes offers a
perceptual interface for exploring the structure of neural signals.
Different parameter regimes produce distinct spectrogram patterns,
which may provide an additional way to investigate the relationship
between entropy measures and neural dynamics.

The framework generates testable predictions. If the entropy–horizon
relation is meaningful, systematic variations in spectral entropy
should modify both the amplitude and spectral structure of the
predicted oscillatory modes. Neural state transitions, such as those
occurring between sleep stages, may correspond to shifts in the
effective accessibility parameter $\Gamma_{r}$.
Conversely, the absence of correlations between entropy variations and the predicted scaling behavior would
falsify the model. The framework therefore provides clear
criteria for empirical validation using large EEG datasets.

Although the present model is conceptual, it provides a
physically motivated framework linking entropy measures,
scale-dependent dynamics, and observable neural oscillations, suggesting new avenues for the quantitative
analysis of complex neural signals.

\emph{Conclusion}--We have introduced a black-hole–inspired horizon  representation of neural signals 
through a complex wavefunction governed by RG dynamics.
Observable neural signals
are treated as boundary-level quantities accessible to external measurement, while internal states remain inaccessible
to direct measurement.
The approach provides a statistical physics perspective on neural signal dynamics and
suggests that observable neural activity may admit an
effective boundary-based description.
Future work may extend the framework to other neural
observables such as magnetoencephalography (MEG) or
large-scale brain-network activity.

\emph{Data availability}--Sonified data illustrating the time evolution of
$\psi$ are available at:
\url{https://youtu.be/s_mPuU6u4JY} and
\url{https://youtu.be/fuwPfcC3xwc}.

\end{document}